\documentclass{iau} 
\usepackage{graphicx}
\usepackage{natbib}

\pubyear{2022}
\volume{373}
\jname{Resolving the Rise and Fall of Star Formation in Galaxies}
\editors{Kim W.-T. \& Wong T., eds.}

\title[SFH of NGC\,5128] 
{Star Formation History of Two Fields in the Halo of NGC\,5128}

\author[Sima T. Aghdam et al]   
{Sima T. Aghdam$^1$, Atefeh Javadi$^1$, Seyedazim Hashemi$^{2,3}$, Jacco Th. van Loon$^4$, Habib Khosroshahi$^1$, Roya H. Golshan$^5$, Elham Saremi$^{1,6,7}$, \and Maryam Saberi$^8$}

\affiliation{$^1$School of Astronomy, Institute for Research in Fundamental Sciences (IPM), Tehran, 19568-36613, Iran \\ email: {\tt sima.t.ahdam@gmail.com} \\[\affilskip]
	$^2$Department of Physics and Astronomy, University of California Riverside, CA 92521, The USA \\
	$^3$Department of Physics, Sharif University of Technology, Tehran, 11155-9161, Iran \\
	$^4$Lennard-Jones Laboratories, Keele University, ST5 5BG, UK \\
	$^5$Physikalisches Institut der Universität zu K{\"o}ln, Zülpicher Str. 77, 50937, Köln, Germany \\
	$^6$Instituto de Astrof{\`i}sica de Canarias, C/ V{\`i}a L{\`a}ctea s/n, 38205 La Laguna, Tenerife, Spain \\
	$^7$Departamento de Astrof{\`i}sica, Universidad de La Laguna, 38205 La Laguna, Tenerife, Spain
	$^8$Rosseland Centre for Solar Physics, University of Oslo, P.O. Box 1029, Blindern, NO-0315, Oslo, Norway}

\begin{document}

\maketitle

\begin{abstract}
NGC\,5128 galaxy is a giant elliptical galaxy located in the Centaurus group of galaxies at $3.8$ Mpc. We aim to study the star formation history (SFH) of two different fields of the galaxy. The northeastern field (Field\,1) is located at a distance of $18.8$ kpc, while the southern field (Field\,2) is at $9.9$ kpc. We use a photometric method that is based on identifying long period variable (LPV) stars and asymptotic giant branch (AGB) stars, as they are strong tracers of star formation and galaxy evolution due to their luminosity and variability; $395$ LPVs in Field\,1 and $671$ LPVs in Field\,2 have been identified. These two fields present similar SFHs, although the SF rate of Field\,2 is more enhanced. We find that the galaxy has three major star formation episodes t $\sim$ $800$ Myr ago, t $\sim$ $3.2$ Gyr ago, and t $\sim$ $10$ Gyr ago, where t is look-back time. The rate of star formation at $\sim$ 800 Myr ago agrees with previous studies suggesting that the galaxy experienced a merger around that time. Furthermore, NGC\,5128 has experienced a lower star formation rate in its recent history which could have been driven by jet-induction star formation and multiple outbursts of AGN activity in this galaxy, as well as a minor merger around $400$ Myr ago.
\keywords{stars: AGB and LPV --
	stars: formation --
	galaxies: jets --
	galaxies: nuclei --
	galaxies: halos --
	galaxies: evolution --
	galaxies: star formation --
	galaxies: individual: NGC\,5128}
\end{abstract}

\firstsection 
              
\section{Introduction} \label{sec:sec1}

There is no giant elliptical galaxy (GE) in the local group (LG), so to study the various features of GE galaxies, we must look at the nearest elliptical galaxy in other groups. NGC\,5128 (or Centaurus A), with a distance of 3.8 Mpc ($\mu =27.87 \pm 0.16 $ mag; \citealp{Rejkuba2004a}, and $E(B-V) = 0.15 \pm 0.05$ mag; \citealp{Rejkuba2001}) provides for us a distinctive opportunity to study the nearest GE galaxy (\citealp{harris1999}; \citealp{charmandaris2000}; \citealp{Rejkuba2004b}; \citealp{Rejkuba2005}), located in the Centaurus group of galaxies (\citealp{karachentsev2005}).

We aim to find the SFH of two small fields in the halo of NGC\,5128 using long period variable stars (LPVs) to better understand the relation between the SFH of the halo and likely recent merger. LPV stars are evolved asymptotic giant branch stars (AGBs) (e.g. \citealp{fraser2005}, \citeyear{fraser2008}; \citealp{soszynski2009}) which are identified according to high luminosity ($1000$ to $60,000$ $L_\odot$), as well as their variability ($\sim$ 100 to 1300 days). In addition, red supergiant (RSG) stars are massive stars with a mass up to $\sim$ $30$ M$_\odot$, known as a tracer of recent star formation (about $10-30$ Myr) in the galaxy with a strong radial pulsation (\citealp{vassiliadis1993}; \citealp{javadi2011b}; \citealp{vanloon1999}, \citeyear{vanloon2005}).

\section{Data} \label{sec:sec2}

We used the data that was obtained by the ISAAC near-IR imaging spectrometer at the ESO Paranal UT1 Antu 8.2m Telescope (VLT) in two different fields in the halo of the galaxy that were published by \cite{Rejkuba2003}, which have been separated into two fields in the halo of the galaxy and identified as Field\,1 and Field\,2 by \cite{Rejkuba2001} and \cite{Rejkuba2003}. Field\,1 is on the distinguished north-eastern part of the halo, at a distance of $\sim$ 18.8 kpc from the center of the galaxy whereas Field\,2 is located at a distance of $\sim$ 9.9 kpc from the center (\citealp{Rejkuba2001}). The detected evolved stars to investigate SFH are the LPV stars with periods longer than $70$ days (from \citealp{Rejkuba2003}). Therefore, $395$ and $671$ LPVs are selected in Field\,1 and Field\,2, respectively.

\section{Method} \label{sec:sec3}

We used a method to study the SFH based on LPVs and it was developed by \cite{javadi2011b} which was explained in the previous papers in details (M33: \citealp{javadi2011b}, \citeyear{javadi2011c}, \citeyear{javadi2016}, \citeyear{javadi2017}; LMC \& SMC: \citealp{Rezaei2014}; NGC\,147 \& NGC\,185: \citealp{golshan2017}; IC\,1613: \citealp{hashemi2019}; Andromeda VII: \citealp{navabi2021}; Andromeda I: \citealp{saremi2021}). In the first step, we should obtain mass, age, and pulsation duration of LPVs (the duration that stars are in the LPV phase) by applying Padova evolutionary models (\citealp{marigo2017}) for various constant metallicities. Furthermore, for studying the star formation history of the galaxy, we consider $Z = 0.039$ (\citealp{woodley2010b}; \citealp{Rejkuba2011}). By assuming all mentioned parameters, the SFRs for different bins with specified intervals in age and mass as follows:

\begin{equation}
\xi(t) = \frac{dn^\prime(t)}{\delta t}\ \frac{\int_{\rm min}^{\rm max}f_{\rm IMF}(m)m\ dm}
{\int_{m(t)}^{m(t+dt)}f_{\rm IMF}(m)\ dm}
\label{eq:eq1}
\end{equation}
where m is mass, $f_{\rm IMF}(m)$ is Kroupa initial mass function (IMF) (\citealp{Kroupa2001}), $dn^\prime$ is the observed LPVs in each bin, and $\delta t$ is the pulsation duration.

The statistical error bars for each bin come from the Poisson distribution:

\begin{equation}
\sigma_{\xi(t)}=\frac{\sqrt{N}}{N}\xi(t)
\label{eq:eq2}
\end{equation}
where N is the number of stars in each age bin.

Stellar evolution in the AGB phase causes that LPV stars inject dust into the interstellar medium (ISM). Therefore, the light from the LPVs will be detected with a fainter magnitude (\citealp{javadi2011b}; \citealp{vanloon1999}, \citeyear{vanloon2005}). To correct the circumstellar extinction, we should do a de-reddened process by plotting the color-magnitude diagram and overlaying the theoretical isochrones by \cite{marigo2017}. 

Furthermore, since NGC\,5128 is a distant galaxy, it is unlikely to detect all of the LPVs in the galaxy. According to \cite{Rejkuba2003} study, they simulated variable stars by using 3 crucial parameters which are the mean magnitude of each star, period, and amplitude. By considering their simulation, we applied the detection probability of each star to our results. It should be noted that it is a key point in the presented method due to its dependency on the number of detected LPVs.

\section{Results and Discussion} \label{sec:sec5}

By considering metallicity Z = $0.039$ and applying the probability functions, the SFHs of Field\,1 and Field\,2 are presented in Fig.\,1. Despite being located in very different parts of the galaxy, these two fields have very similar SFHs, as well as the higher rates of star formation for Field\,2. There is a similar supply of gas to stars during their evolution. In each field, the consistent patterns of SFRs is seen that are comparable for Field\,1 and Field\,2 in increasing in log\,t(yr) $\sim$ $8.9$ (t $\sim$ $800$ Myr), log\,t(yr) $\sim$ $9.5$ (t $\sim$ $3.2$ Gyr), and log\,t(yr) $\sim$ $10$ (t $\sim$ $10$ Gyr) where $t$ is look-back time.

\begin{figure}[h]
\begin{center}
	\includegraphics[width=0.8\textwidth]{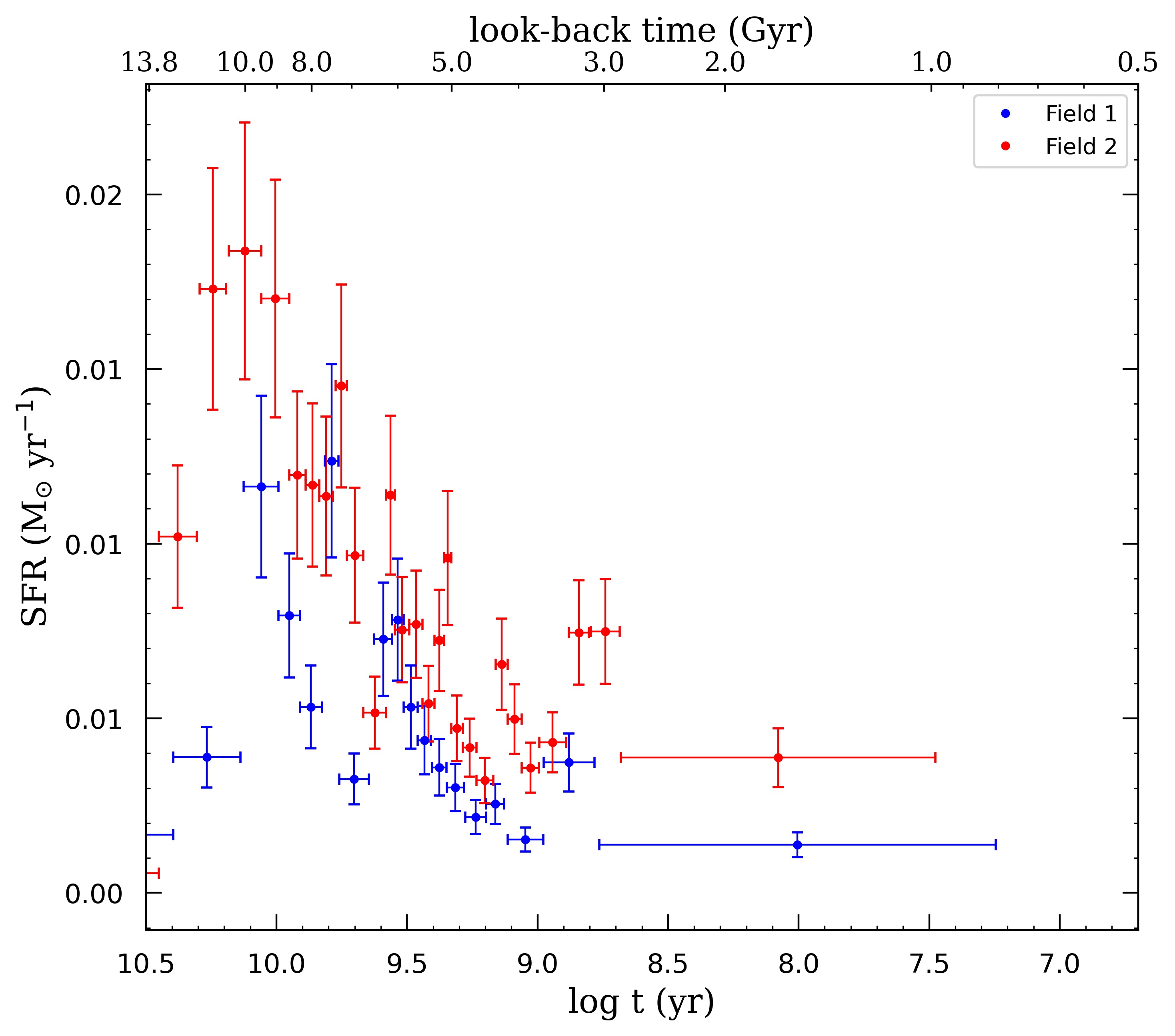}
	\caption{The SFH of Field\,1 and Field\,2 for Z = $0.039$. {\it The blue} represents the SFH of Field\,1, while {\it the red} one indicates the SFH of Field\,2.}
	\label{fig:fig1}
\end{center}
\end{figure}

Additionally, an increase in star formation in Field\,1 and Field\,2 around $800$ Myr ago is consistent with the idea of recent merger of NGC\,5128 with a small gas-rich galaxy (\citealp{israel1998}; \citealp{fassett2000}; \citealp{mould2000}; \citealp{Rejkuba2001}). After log\,t(yr) $\sim$ $8.5$, the rate of recent star formation stays constant according to the recent merger which happened around $400$ Myr ago (\citealp{peng2002}). Furthermore, NGC\,5128 has experienced AGN activity based on the central supermassive black hole (SMBH). There is an acceptable scenario leading to starbursts and AGN activity which is the interaction of gas-rich galaxies during mergers (\citealp{mo2010}).

\def\apj{{ApJ}}    
\def\nat{{Nature}}    
\def\jgr{{JGR}}    
\def\apjl{{ApJ Letters}}    
\def\aap{{A\&A}}   
\def\mnras{{MNRAS}}
\def\aj{{AJ}}
\let\mnrasl=\mnras

\end{document}